\documentclass[useAMS,usenatbib]{mn2e}
\usepackage{graphicx}
\usepackage{txfonts}
\title[SAX J2103.5+4545]{Correlated optical/X-ray variability in the high-mass X-ray
   binary SAX J2103.5+4545}
\author[P. Reig et al.]{P. Reig$^{1,2}$\thanks{E-mail: pau@physics.uoc.gr},
    A. S\l{}owikowska$^{1,2}$, 
	A. Zezas$^{1,2,3}$ and P. Blay$^{4}$
\\
$^{1}$IESL, Foundation for Research and Technology, 71110 Heraklion,
Crete, Greece\\
$^{2}$University of Crete, Physics Department, PO Box 2208, 710 03
Heraklion, Crete, Greece \\
$^{3}$ Harvard-Smithsonian Center for Astrophysics, 60, Garden Street,
Cambridge, MA 02138, USA \\
$^{4}$Image Processing Laboratory, University of Valencia, 46071 Paterna-Valencia, Spain 
}

\newcommand{\src}  {SAX J2103.5+4545}
\newcommand{\ha}  {H$\alpha$}

\newcommand{\ew}     {EW(H$\alpha$)}

\def\simless{\mathbin{\lower 3pt\hbox
     {$\rlap{\raise 5pt\hbox{$\char'074$}}\mathchar"7218$}}}   
\def\simmore{\mathbin{\lower 3pt\hbox
     {$\rlap{\raise 5pt\hbox{$\char'076$}}\mathchar"7218$}}}   

\def\msun{~{\rm M}_\odot}
\def\rsun{~{\rm R}_\odot}

\begin{document}

\date{Accepted ??. Received ??; in original form ??}

\pagerange{\pageref{firstpage}--\pageref{lastpage}} \pubyear{2009}

\maketitle

\label{firstpage}

\begin{abstract}

SAX J2103.5+4545 is the Be/X-ray binary with the shortest orbital period.
It shows extended bright and faint X-ray states that last for a few hundred
days. The main objective of this work is to investigate the relationship
between  the X-ray and optical variability and to characterise the spectral
and timing  properties of the bright and faint states. We have found a
correlation between the spectral and temporal parameters that fit the
energy and power spectra. Softer energy spectra correspond to softer power
spectra. That is to say, when the energy spectrum is soft the power at high
frequencies is suppressed. We also present the results of our monitoring of
the H$\alpha$ line of the optical counterpart  since its discovery in 2003.
There is a correlation between the strength and shape of the H$\alpha$
line, originated in the circumstellar envelope  of the massive
companion and the X-ray emission from the vicinity of the neutron  star.
H$\alpha$ emission, indicative of an equatorial disc around the B-type
star, is detected  whenever the source is bright in X-rays. When the disc
is absent, the X-ray  emission decreases significantly. The long-term
variability of \src\ is  characterised by fast episodes of disc loss and
subsequent reformation. The time scales for the loss and reformation of the
disc (about 2 years) are the fastest among Be/X-ray binaries. 

\end{abstract}


\begin{table}[h]
\begin{center}
\caption{H$\alpha$ equivalent width measurements ($1\sigma$
errors). See Fig.~\ref{asmha} for a representative example of the line profiles.}
\label{ewha}
\begin{tabular}{ccccc}
\hline \hline \noalign{\smallskip}  
Date	&Julian date	&EW(\ha)	&\ha		&Telescope	\\
	&(2,400,000+)	&(\AA)		&profile$^a$	&		\\
\hline \noalign{\smallskip}
01/08/2003 &52853.5	&-2.14$\pm$0.16	&DPE	&SKI	\\
17/08/2003 &52869.6	&-1.06$\pm$0.09	&SH	&WHT	\\
14/09/2003 &52897.3	&+2.51$\pm$0.11	&ABS	&WHT	\\
06/10/2003 &52919.3	&+2.31$\pm$0.16	&ABS	&SKI	\\
08/10/2003 &52921.3	&+2.08$\pm$0.12	&CQE	&SKI	\\
03/12/2003 &52977.3	&+2.11$\pm$0.31	&ABS	&CAL	\\
23/05/2004 &53149.6	&+2.61$\pm$0.24	&CQE	&SKI	\\
27/05/2004 &53153.6	&+2.61$\pm$0.34 &CQE    &SKI    \\
28/05/2004 &53154.5	&+2.50$\pm$0.19 &CQE	&SKI	\\
23/06/2003 &53180.5	&+1.90$\pm$0.08 &CQE	&SKI	\\
25/06/2004 &53182.5	&+2.39$\pm$0.27 &ABS	&SKI	\\
07/07/2004 &53194.5	&+2.80$\pm$0.17 &ABS	&SKI	\\
25/08/2004 &53243.5	&+2.34$\pm$0.15 &CQE	&SKI	\\
26/08/2004 &53244.4	&+2.15$\pm$0.12 &CQE	&SKI	\\
27/08/2004 &53245.4	&+1.93$\pm$0.10 &CQE	&SKI	\\
01/09/2004 &53250.6	&+2.33$\pm$0.11 &ABS	&WHT	\\
03/09/2004 &53252.5	&+2.30$\pm$0.13 &ABS	&SKI	\\
12/09/3004 &53261.4	&+2.16$\pm$0.07 &ABS	&SKI	\\
13/09/2004 &53262.3	&+2.11$\pm$0.18 &ABS	&SKI	\\
25/10/2004 &53304.3	&+2.49$\pm$0.31 &CQE	&SKI	\\
23/06/2005 &53545.4	&+2.31$\pm$0.13 &CQE	&SKI	\\
11/07/2005 &53563.5	&+2.59$\pm$0.11 &ABS	&SKI	\\
29/07/2005 &53581.4	&+2.46$\pm$0.10 &ABS	&SKI	\\
16/08/2005 &53599.5	&+2.27$\pm$0.08 &ABS	&SKI	\\
20/09/2005 &53634.4	&+2.35$\pm$0.06 &ABS	&SKI	\\
26/10/2005 &53670.4	&+1.21$\pm$0.05 &CQE	&SKI	\\
20/06/2006 &53907.5	&+2.43$\pm$0.11 &ABS	&SKI	\\
03/10/2006 &54012.3	&+1.98$\pm$0.12 &CQE	&SKI	 \\
14/05/2007 &54235.4	&-2.29$\pm$0.24 &DPE	&SKI	 \\
20/05/2007 &54241.4	&-1.72$\pm$0.21 &DPE	&SKI	 \\
29/05/2007 &54250.5	&-1.58$\pm$0.09 &DPE	&SKI	 \\
09/06/2007 &54261.9	&-1.19$\pm$0.11 &DPE	&FLW   \\
22/06/2007 &54274.9	&-1.06$\pm$0.14 &DPE	&FLW   \\
04/09/2007 &54348.5	&-5.02$\pm$0.31 &DPE	&SKI	 \\
06/09/2007 &54350.5	&-4.18$\pm$0.32 &DPE	&SKI	 \\
09/09/2007 &54353.5	&-5.12$\pm$0.41 &DPE	&SKI	 \\
11/09/2007 &54355.4	&-4.88$\pm$0.24 &DPE	&SKI	 \\
14/09/2007 &54358.9	&-4.35$\pm$0.28 &DPE	&FLW   \\
02/10/2007 &54376.3	&-4.65$\pm$0.36 &DPE	&SKI	 \\
03/10/2007 &54377.3	&-4.24$\pm$0.41	&DPE	&SKI	\\
05/12/2007 &54439.6	&-1.00$\pm$0.10 &SH	&FWL  \\
24/06/2008 &54642.4	&+2.02$\pm$0.09 &CQE	&SKI	\\
25/06/2008 &54643.4	&+2.18$\pm$0.14 &CQE	&SKI	\\
14/07/2008 &54662.4	&+2.21$\pm$0.10 &CQE	&SKI	\\
08/08/2008 &54687.4	&+2.33$\pm$0.08 &CQE	&SKI	\\
12/08/2008 &54691.3	&+1.93$\pm$0.17 &ABS	&SKI	\\
02/09/2008 &54712.3	&+2.26$\pm$0.04 &CQE	&SKI	 \\
08/05/2009 &54960.5	&+2.10$\pm$0.13 &CQE	&SKI	\\
17/05/2009 &54969.5	&+2.03$\pm$0.13 &ABS	&SKI	\\
27/05/2009 &54979.6	&+2.16$\pm$0.11 &ABS	&SKI	\\
\hline \hline \noalign{\smallskip}
\multicolumn{5}{l}{$a$: ABS: absorption, CQE: absorption with central quasi-emission peak}\\
\multicolumn{5}{l}{\, \, \, DPE: double-peak emission, SH: shell profile. }\\
\end{tabular}
\end{center}
\end{table}

\section{Introduction}

High-mass X-ray binaries (HMXB) typically consist of a neutron star orbiting around
an early-type star. They divide into two major groups according to the
luminosity class of the optical component: supergiant X-ray binaries if the
massive companion is an evolved star (luminosity class I-II) and Be/X-ray
binaries (BeX) if it is a main-sequence or giant star. The vast majority of
BeX are transient systems, although there is a small group of persistent
sources \citep{reig99}. Transient BeX have orbital periods in the range
20-100 days and spin periods shorter than  $\sim$200 s. In contrast,
persistent BeX present orbital periods typically longer than $\sim$100 days
and spin periods longer than $\sim$200 s. Supergiant X-ray binary systems
have narrower orbits with orbital periods shorter than 10 days and contain
slow-rotating neutron stars, with spin periods of the order of minutes to
hours.

\src\ is an unusual HMXB. Its spin, $P_{\rm spin}= 358.6$ s \citep{hull98},
and orbital, $P_{\rm orb}=12.7$ d \citep{bayk02} periods are typical of
supergiant X-ray binaries. In fact, \src\ falls in the wind-fed supergiant
region \citep{reig04} of the $P_{\rm orb}-P_{\rm spin}$ diagram
\citep{corb86}. However, the primary component of the binary is a
main-sequence star of spectral type B0 \citep{reig04}, i.e. \src\ belongs
to the BeX category.  \src\ alternates X-ray bright states that last for a
few months with extended (as long as few years) X-ray faint states \citep{bayk02}. 
During the bright states \src\ shows large spin-up episodes
\citep{inam04,sido05}, indicating that an accretion disc is formed around
the neutron star \citep{bayk02}. The scarce optical data seem to
suggest that during the bright state the H$\alpha$ line turns into
emission, implying that a decretion circumstellar disc is formed around the
Be star \citep{reig04}. These discs are probably short lived and appear
during the high X-ray emission states only. \citet{reig05} found that \src\
was emitting X-rays even after the complete loss of the circumstellar disc.
They argue that in this state the X-rays are the result of wind-fed
accretion. 

Since the discovery of \src\ in February 1997, the system has been observed in
the X-ray band on numerous occasions. The first X-ray observations were made
with {\it BeppoSAX} \citep{hull98}. The source was active for about eight months
and reached a peak intensity of 20 mCrab (2--25 keV) on April 11, 1997. The
X-ray emission showed pulsations with a period of 358.61 s. 

The All Sky Monitor on board {\it RXTE} detected a second bright state two years
later \citep{bayk00}, reaching a peak intensity of 27 mCrab (2-12 keV) on
October 28, 1999. The continuous monitoring of \src\ by {\it RXTE} and {\it
INTEGRAL} since July 2002 allowed  detailed pulse frequency analysis and the
determination of the orbital parameters \citep{bayk07,came07}: the system has a
moderately eccentric orbit with $e=0.401\pm0.018$,  an orbital period of
$12.66528\pm0.00051$ days, and a semi-major axis of $80.8\pm0.7$ lt-s.

In addition to {\it BeppoSAX} and {\it RXTE}, \src\ has been observed with
{\it XMM-Newton} and {\it INTEGRAL} in the bright state, hence
allowing the characterisation of the energy spectrum from 0.1 to 150 keV.
At low energies ($<$ 4 keV), the X-ray energy spectrum presents a soft
spectral component, consistent with blackbody emission with $kT=1.9$ keV
and emitting radius of $\sim$0.3 km \citep{inam04}. The broad-band (4-150
keV) energy spectrum is well fitted by a power-law ($\Gamma=1-1.5$) with an
exponential cutoff ($E_{\rm cut}=8-19$ keV), and with a K$\alpha$
fluorescence emission line ($\sim$6.4 keV) from cool iron
\citep{blay04,sido05}. 

The spectral analysis of RXTE data performed by \citet{bayk02} covering the
interval November 1999--August 2000 indicate that the spectrum of \src\ is harder
during bright states. This result was confirmed by the X-ray colour analysis
performed by \citet{came07} using six years worth of RXTE data.
Transient quasi-periodic oscillations around 0.044 Hz have been reported by
\citet{inam04}.

The excellent monitoring of \src\ in the X-ray band contrasts with the
scarcity of data at other wavelengths. The only published optical/near-IR
observations of \src\ are those reported by \citet{reig04}. They identified
the optical counterpart with a moderately reddened V=14.2
B0V star, at a distance of 6.5 kpc. In this paper we present the results of
our monitoring of the H$\alpha$ line for the 2003-2009 interval and
investigate the X-ray variability of \src\ in correlation with the optical
data. X-rays result from accretion of matter from the optical companion's
circumstellar disc onto the neutron star surface, hence providing
information about the physical conditions in the vicinity the compact
object. The \ha\ line is formed in the Be star's disc as
reprocessed radiation from its photosphere. Since the disc constitutes the
reservoir of matter available for accretion a correlated X-ray/optical
study is the best way to unveil the nature of the variability in this system.

\begin{table*}
\begin{center}
\caption{Optical photometric observations of \src.}
\label{phot}
\begin{tabular}{lccccc}
\hline \hline \noalign{\smallskip}
Date		&Julian Date	&B		&V		&R	&I  \\
		&2,400,000+	&mag		&mag		&mag	&mag \\
\hline \noalign{\smallskip}
08-06-2003 &52799.467 &15.35$\pm$0.03 &14.22$\pm$0.02 &13.48$\pm$0.02 &--   \\
24-08-2003 &52876.408 &15.36$\pm$0.02 &14.25$\pm$0.03 &13.57$\pm$0.03 &12.85$\pm$0.03   \\
20-05-2004 &53146.558 &15.41$\pm$0.01 &14.32$\pm$0.01 &13.63$\pm$0.01 &12.87$\pm$0.01   \\
05-07-2004 &53192.354 &15.42$\pm$0.02 &14.33$\pm$0.01 &13.64$\pm$0.02 &12.92$\pm$0.02   \\
24-08-2004 &53242.477 &15.39$\pm$0.02 &14.31$\pm$0.01 &13.61$\pm$0.01 &12.85$\pm$0.02   \\
14-09-2004 &53263.375 &15.39$\pm$0.02 &14.27$\pm$0.02 &13.58$\pm$0.01 &12.84$\pm$0.03   \\
01-10-2004 &53280.370 &15.41$\pm$0.02 &14.31$\pm$0.02 &13.61$\pm$0.02 &12.86$\pm$0.02   \\
26-06-2005 &53548.497 &15.39$\pm$0.02 &14.30$\pm$0.01 &13.61$\pm$0.02 &12.86$\pm$0.04   \\
27-07-2005 &53579.502 &15.38$\pm$0.02 &14.29$\pm$0.01 &13.60$\pm$0.01 &--   \\
20-08-2005 &53603.414 &15.42$\pm$0.02 &14.35$\pm$0.03 &13.66$\pm$0.02 &--   \\
18-08-2006 &53966.519 &15.42$\pm$0.03 &14.32$\pm$0.03 &13.63$\pm$0.02 &--   \\
16-07-2007 &54298.493 &15.12$\pm$0.03 &13.75$\pm$0.03 &12.93$\pm$0.03 &12.05$\pm$0.03   \\
01-09-2007 &54345.426 &15.21$\pm$0.03 &13.94$\pm$0.02 &13.15$\pm$0.03 &12.32$\pm$0.04   \\
02-09-2007 &54346.496 &15.12$\pm$0.03 &13.89$\pm$0.03 &13.11$\pm$0.02 &12.28$\pm$0.02   \\
26-10-2007 &54400.242 &15.37$\pm$0.03 &14.21$\pm$0.03 &13.57$\pm$0.04 &12.78$\pm$0.05   \\
05-08-2008 &54684.434 &15.41$\pm$0.02 &14.29$\pm$0.02 &13.60$\pm$0.02 &12.87$\pm$0.02   \\
\noalign{\smallskip} \hline \hline 
\end{tabular}
\end{center}
\end{table*}

\section{Observations and data analysis}

\subsection{Optical spectra}

We have been observing \src\ regularly  since the identification of the
optical counterpart in 2003. We have analysed all available optical
observations including those already published by \citet{reig04}.  The new
optical spectroscopic observations were obtained mainly from two sites: the
Skinakas (SKI) observatory in Crete (Greece) and from the Fred Lawrence
Whipple (FLW) observatory  at Mt. Hopkins (Arizona). In addition, \src\ was
observed in service time from the William Herschel Telescope (WHT) at the
observatory of El Roque de los Muchachos in La Palma (Spain) on 1 September
2004. One more spectrum was made from the 2.2-m telescope at the
observatory of Calar Alto (CAL) in Almer\'{\i}a (Spain) on 3 December 2003.
Table~\ref{ewha} gives the log of the spectroscopic observations and the
value of the H$\alpha$ equivalent width.

The 1.3\,m telescope of the Skinakas Observatory was equipped with a
2000$\times$800 (15 $\mu$m) pixel ISA SITe CCD and a 1302 l~mm$^{-1}$ grating, giving a
nominal dispersion of $\sim$1 \AA/pixel.  The instrumental set-up during
the service WHT observation consisted of  the ISIS  spectrograph with the
R1200R grating, giving a dispersion of 0.23 \AA/pixel and covering the
wavelength interval between 6040--6900 \AA. The spectrum was obtained with
the  4096$\times$2048 (13.5 $\mu$m) pixels  MARCONI2 CCD.  The reduction of
these spectra was made using the STARLINK {\em Figaro} and IRAF v2.14
packages, while their analysis was performed using the STARLINK {\em Dipso}
package. The FLW observations of \src\ were made in queue mode with the
1.5-m  telescope and the FAST-II spectrograph \citep{fabr98}.  Three
observations were obtained with the 1200 l~mm$^{-1}$ grating and one
observation (5 December 2007) with the 600 l~mm$^{-1}$ grating.  The tilt
of the grating was selected in order to cover the $\rm{H\alpha}$ line. The
data were analysed with the RoadRunner package \citep{toka97} implemented
in IRAF.  Spectra of comparison lamps were taken before each exposure in
order to account for small variations of the wavelength calibration during
the night.

The field around \src\ was observed through the Johnson $B$, $V$, $R$, and
$I$ filters from the 1.3\,m telescope of the Skinakas observatory. 
Observations obtained before July 2007 were performed using a $1024 \times
1024$ SITe CCD chip with a 24 $\mu$m pixel size (corresponding to
$0.5^{\prime\prime}$ on sky). From July 2007 a 2048$\times$2048 ANDOR CCD
with a 13.5 $\mu$m pixel size was used (corresponding to
$0.28^{\prime\prime}$ on sky). Standard stars from the Landolt list
\citep{land09} were used for the photometric calibrations and
transformation equations.  Reduction of the data was carried out in the
standard way using the IRAF tools for aperture photometry. 

\begin{table}
\begin{center}
\caption{Summary of the RXTE observations.}
\label{xobs}
\begin{tabular}{ccccc}
\hline \hline \noalign{\smallskip}
Proposal &MJD		&On-source	&$<I_{\rm x}>^a$	&$rms^b$ \\
ID	 &start		&time (ks)	&cs$^{-1}$		&(\%)\\
\hline \noalign{\smallskip}
P40125	&51501.1846	&148.05		&20.1	&54	\\
P40438	&51559.9448	&89.79		&24.3	&57	\\
P50095	&51606.2713	&154.16		&21.4	&79	\\
P50417	&51764.0044	&60.35		&7.0	&40	\\
P60409	&52019.0087	&5.20		&24.1	&41	\\
P70082	&52439.9567	&446.51		&33.2	&71	\\
P90097	&53029.1948	&252.86		&9.94	&67	\\
P92436	&54221.0628	&27.98		&40.54	&86	\\
\hline \hline \noalign{\smallskip}
\multicolumn{5}{l}{$a$: 2-60 keV PCU2 background-subtracted. Time bin: 16 s}\\
\multicolumn{5}{l}{$b$: root-mean-square: standard deviation over the mean
intensity}\\
\end{tabular}
\end{center}
\end{table}

\subsection{X/$\gamma$-ray observations}

We have analysed all archived observations of \src\ made by {\it RXTE}/PCA
and {\it INTEGRAL}/ISGRI. The {\it RXTE}/PCA consists of five Proportional
Counter Units (PCU), which are sensitive to X-rays in the 2-60 keV energy
range \citep{jaho96}. The INTEGRAL Soft Gamma Ray Imager (ISGRI) is the
lower energy detector of the Imager on Board INTEGRAL Satellite (IBIS) and
operates in the 15--1000~keV energy range \citep{lebr03}.

\begin{figure}
\resizebox{\hsize}{!}{\includegraphics{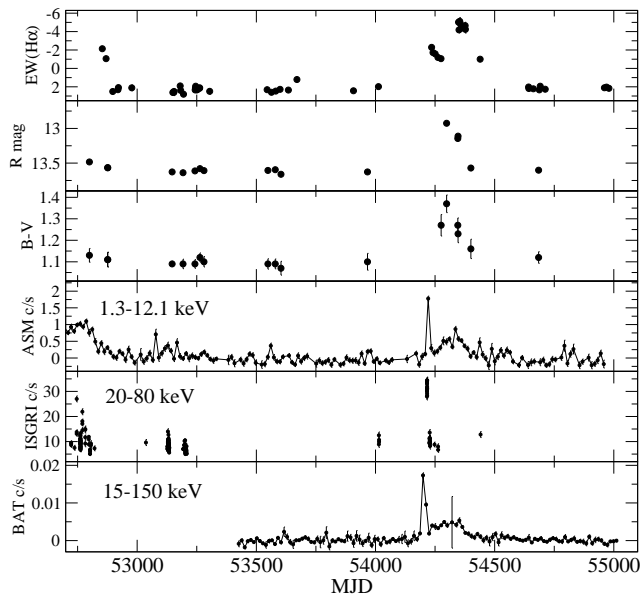} } 
\caption[]{Long-term evolution of the H$\alpha$ equivalent width, R band
and $B-V$ colour. Also shown is the long-term variability in three X-ray
bands, corresponding to the 
{\it RXTE}/ASM, {\it INTEGRAL}/ISGRI and {\it SWIFT}/BAT light curves.  }
\label{xewR}
\end{figure}

In addition, data from two all sky monitors, namely the  All-Sky Monitor
(ASM) onboard {\it RXTE} \citep{levi96} and the Burst Alert Telescope (BAT)
onboard {\it SWIFT} \citep{bart00} have been used to study the long-term
variability of \src\ (Fig.~\ref{xewR}). These two instruments produce daily
flux averages in the energy range 1.3-12.1 keV and  15-150 keV,
respectively.

{\it RXTE}/PCA data can be collected and telemetered to the ground in many
different ways depending on the intensity of the source and the spectral
and timing resolution desired. In this work we used the two standard modes:
{\em Standard1} provides 0.125-s resolution and no energy resolution; in
the {\em Standard2} configuration data are accumulated every 16 seconds in
129 channels.  For the timing analysis {\em GoodXenon} data, providing
maximum spectral (256 channels) and temporal ($2^{-20}$ s) resolution, have
also been employed. The number of active proportional counter units varied
during the observations.  However, PCU2 was always on. To avoid
complications with the different response of the each unit and the varying
number of active detectors we used PCU2 data for the spectral analysis.
Note that the power spectral analysis is not affected by the number of
active units as we used the $rms$ normalisation \citep{bell90,miya91},
where the power is normalised to the mean intensity of short (128 s)
segments. Table~\ref{xobs} gives a log of the {\it RXTE} observations. 

A total of 3513  public data pointings of INTEGRAL included \src\ in the
IBIS/ISGRI field of view. However, only in 269 the source was detected with
significance greater than 3$\sigma$ above the noise level (detection level
greater than 7 based on the standard Offline Scientific Analysis (OSA)
software\footnote{The OSA software generates sky images and searches for
significant sources. A source is considered as detected if it has a
detection level greater than 7, which is equivalent to the classical
3$\sigma$ level above the measured noise.}), while in 1436 additional
observations we obtained marginal or non reliable detections. In the
remaining 1808 pointings the source was below the  sensitivity level of
IBIS/ISGRI. 

OSA version 7.0 have been used to perform light curve and spectra
extraction for each science window. To obtain average spectra corresponding
to several science windows we used an alternative event extraction method
developed by \citet{ferr07}.

\begin{figure}
\resizebox{\hsize}{!}{\includegraphics{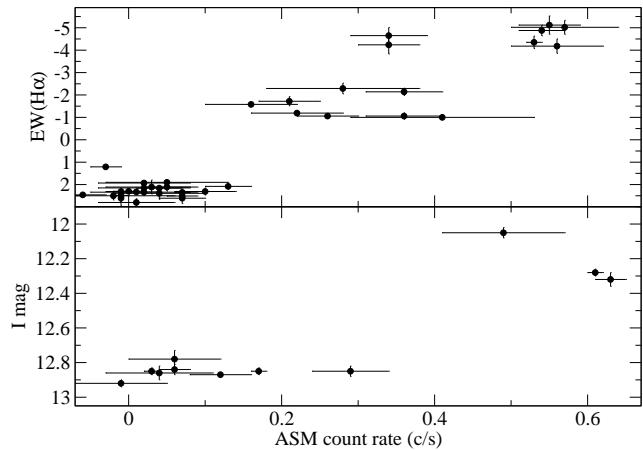} } 
\caption[]{\ew\ and I-band magnitude as a function of ASM intensity.  }
\label{xopt}
\end{figure}

\begin{figure*}
\resizebox{\hsize}{!}{\includegraphics{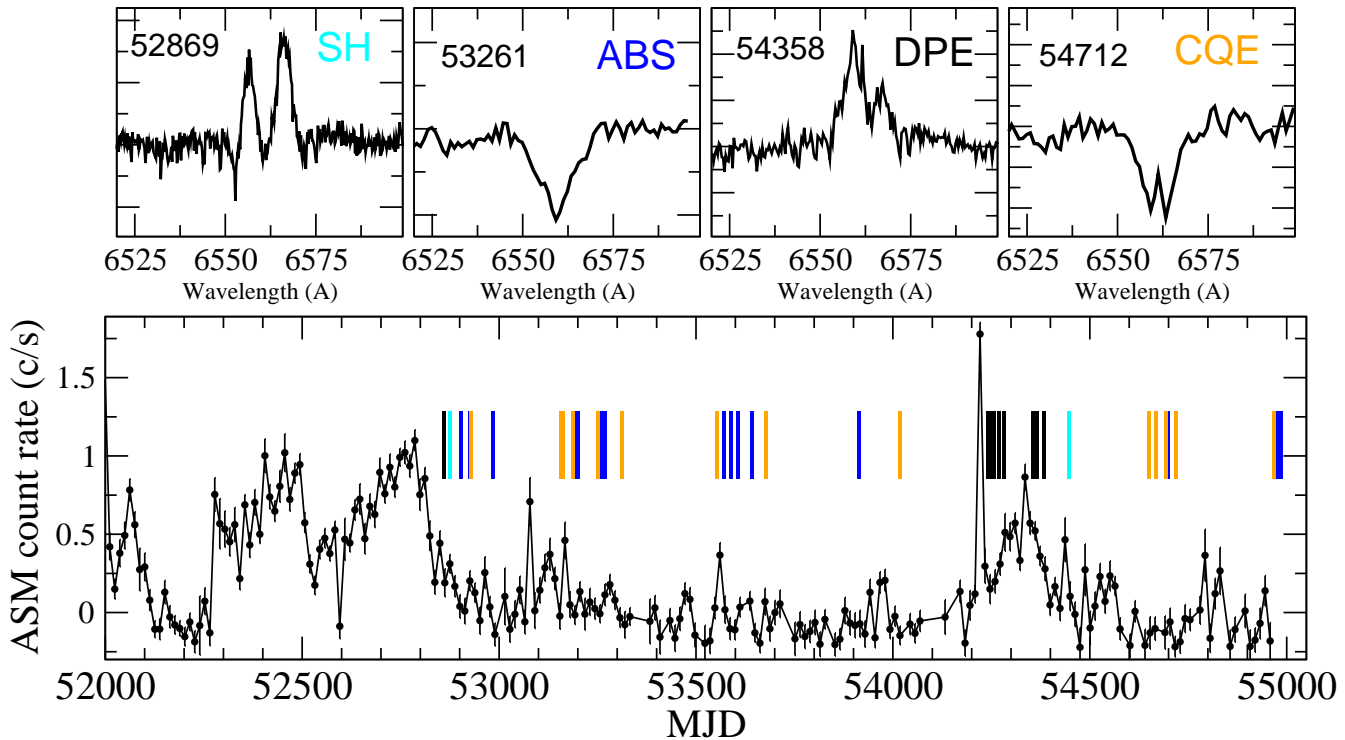}} 
\caption[]{H$\alpha$ line profiles at different X-ray activity states. 
{\it black}: emission (DPE), {\it blue}: absorption (ABS), {\it orange}: central
quasi-emission (CQE) and {\it cyan}: shell profiles (SH).
H$\alpha$ emission occurs when the source is X-ray active. {\em See the 
electronic edition of the Journal for a colour version of 
this figure}}.
\label{asmha}
\end{figure*}

\section{Results}

\subsection{Optical photometry}

Figure~\ref{xewR} shows the evolution of the \ha\ equivalent width (\ew),
R-band magnitude and the $B-V$ colour together with the long-term X-ray
variability in three different energy bands. Despite the observational gaps
in the optical data a correspondence between the variability in the optical
band and the X-ray intensity is apparent. Bright X-ray states correspond to
enhanced optical activity, both in the photometric magnitudes and colours
and the strength of the \ha.   This correlation is more clearly seen
in Fig.~\ref{xopt}, where the \ew\ and I-band magnitude as a function of
ASM intensity is displayed. 

As expected \citep[][and references therein]{doug94}, the contribution of
the Be star's disc to the photometric magnitudes increases with wavelength.
As can be deduced from Table~\ref{phot}, the amplitude of variability of
the optical magnitudes between maximum and minimum brightness is 0.3 mag in
the B band, 0.6 mag in the V band, 0.7 mag in the R band and 0.8 mag in the
I band. 

In \citet{reig04} values of the reddening and distance were derived from the
spectra and the photometric observations. Since {\em i)} a much longer
set of optical photometric observations is now available and {\em ii)} a longer
extended low optical state is observed, where presumably the
underlying B star is exposed and photospheric emission without substantial 
contribution from the disc is detected, it is justified to revisit these values.
The main source of uncertainty in estimating the distance stems from the
uncertainty of the intrinsic colours and absolute magnitudes associated to each
spectral and luminosity class. Difference between different calibrations may
amount up to 1.5 mag in the values of the absolute magnitude \citep{wegn06},
while differences of 0.04 mag in intrinsic $(B-V)_0$ are found in studies from
different authors \citep{john66,wegn94}. \citet{jasc98} analysed the absolute
magnitude of about one hundred MK standards and concluded that the intrinsic
dispersion of the mean absolute magnitude amounts to 0.7 mag. We have adopted
this value as the error on the absolute magnitude.

To estimate the distance to \src\ we use the observations showing the
bluest colours (those from 20 August 2005, MJD 53603), as no contribution
from the circumstellar disc is expected. By comparing the observed colour
$(B-V)=1.07\pm0.03$ with the expected one of a B0V star $(B-V)_0=-0.28$
\citep{john66,guti79,wegn94}, we derive a colour excess of
$E(B-V)=1.35\pm0.03$.  Taking the standard reddening law $A_V=(3.1\pm0.1)
E(B-V)$ and assuming an average absolute magnitude for a B0V star of
$M_V=-4.0\pm0.7$ \citep{vacc96,wegn06} the distance to \src\ is estimated
to be $\sim$ 6.8$\pm$2.3 kpc. The final error was obtained by propagating
the errors of $(B-V)$ (0.03 amg), $A_V$ (0.17 mag) and $M_V$ (0.7 mag). The
estimated distance is consistent with the value of 4.5$\pm$0.5 kpc derived
from the relationship between the pulse frequency derivative and the X-ray
flux \citep{bayk07}.

\subsection{\ha\ line profiles}

The monitoring of \src\ in the optical band reveals that the \ha\ line is
highly variable, both in strength and shape. In Be stars, the strength and
shape of the H$\alpha$ line provide information about the physical
conditions in the circumstellar disc. The measurements of the \ew\ are
given in Table~\ref{ewha}. A negative value indicates that the line appears
in emission. The fourth column of Table~\ref{ewha} indicates the profile of
the line. We have observed four different profiles (see Fig.~\ref{asmha}):
emission with a shell profile (SH), absorption (ABS), emission with a split
profile (DPE) and central (quasi)-emission peak (CQE). The difference
between double-peak profile and a shell profile is simply that in the
latter the depression between the two peaks extends below the stellar
continuum. CQE is a type of absorption line profile in which the central
part of the line core exhibits a weak local flux maximum \citep{rivi99}.

What determines whether the \ha\ line appears in absorption or emission is
the presence of an equatorially concentrated circumstellar disc around the
Be star. The absorption lines are formed in the photosphere of the star
(i.e., no disc is present) while the emission profile is the result of
recombination radiation from ionised hydrogen in the hot, extended
circumstellar envelope surrounding the central Be star. CQEs appear when
the innermost regions of the disc are being supplied with matter. The deep
central absorption in a shell profile is merely due to self-absorption in
the envelope as a consequence of a high inclination angle \citep[see
e.g.][]{hanu95}. CQE and shell features are related phenomena. CQEs occur
only when shell lines are also present and the lines showing CQE profiles
are those also showing a shell feature. Figure~\ref{asmha} show a
representative example of each one of these line profiles.

During the bright states, the \ha\ line exhibits an emission
profile, indicating the presence of a circumstellar disc, whereas during
the faint states the source shows a purely photospheric absorption
profile or a CQE-type profile.

\begin{figure*}
\resizebox{\hsize}{!}{\includegraphics{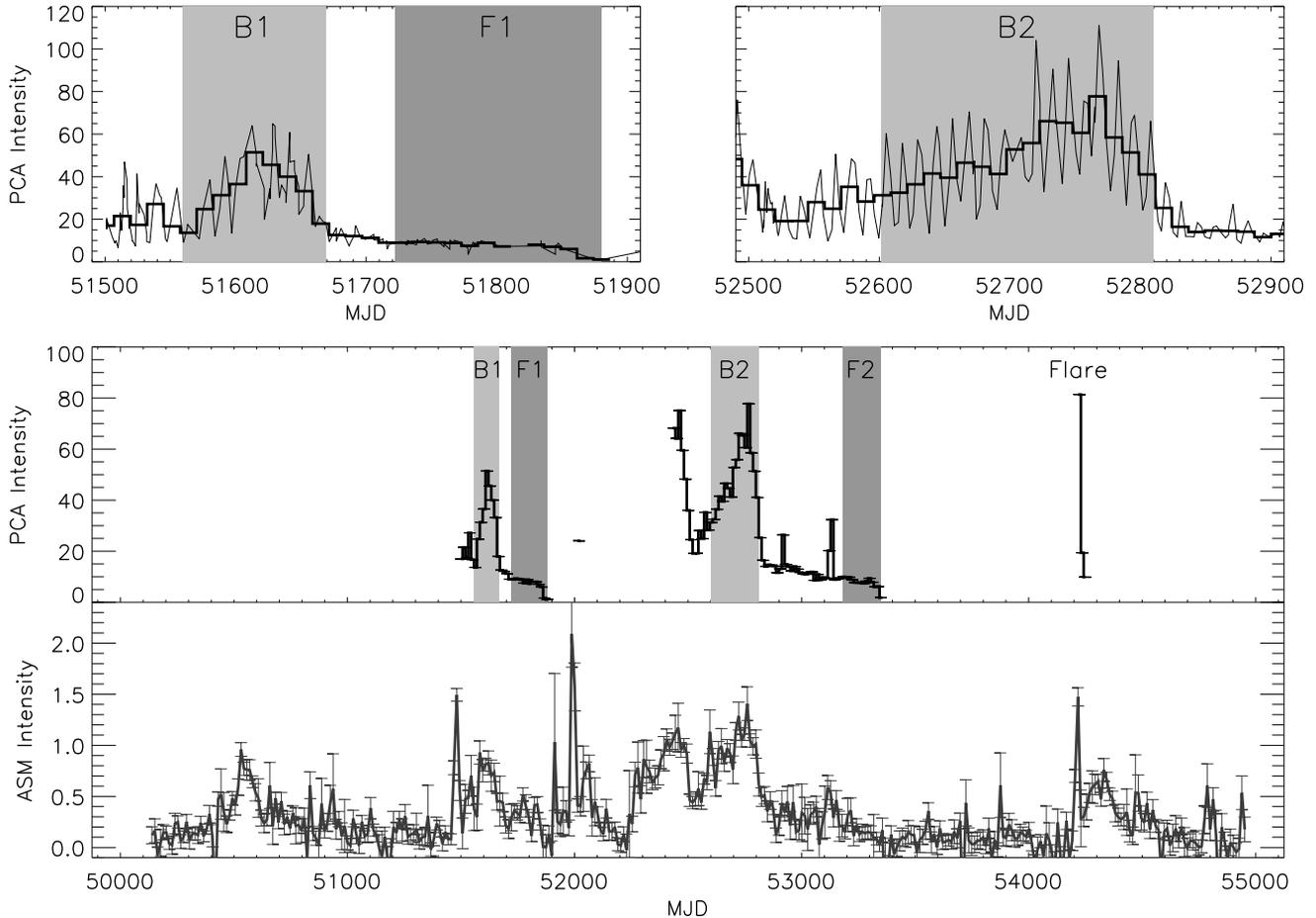} } 
\caption[]{{\it Bottom panel}: {\it RXTE}/ASM light curve (1.3-12.1 keV) since 
April 1996. A bin size equal to the orbital period, $P_{\rm orb}=12.67$ d is 
used.
{\it Middle panel}: {\it RXTE}/PCU2 (2-60 keV) light curve showing 
bright (B1, B2) and faint states (F1, F2). {\em Top panel}: Detailed view of the PCU2 light
curve. The 12.7-d binned PCA light curves is superposed on the average count rate
per observation interval (typically, a few thousand seconds integration time).
Note the modulation of the X-ray 
intensity with the orbital period during the bright states and the lack of
it during the faint state. }
\label{asmpca}
\end{figure*}

\begin{table*}
\begin{center}
\caption{Spectral parameters resulting from the fits to the {\it RXTE}/PCA alone and
{\it RXTE}/PCA+{\it INTEGRAL}/ISGRI energy spectra. Errors denote 90\% confidence.}
\label{pcafit}
\begin{tabular}{lcccccc}
\hline \hline \noalign{\smallskip}
Parameter		&Bright1		&Bright2	&Faint1		&Faint2			&Flare			\\
			&51559--51669		&52602--52810	&51722--51880	&53180--53348		&54221\\
\hline \noalign{\smallskip}
\multicolumn{6}{c}{{\it RXTE}/PCA energy spectra}\\
\hline \noalign{\smallskip}
$N_{\rm H}^a$		&3.7$\pm$0.5		&3.7$\pm$0.4	&4.6$\pm$0.9	&3.8$\pm$0.8		&2$\pm$1		\\
Photon index		&0.85$\pm$0.05		&0.81$\pm$0.02	&1.26$\pm$0.06	&1.21$\pm$0.06		&0.7$\pm$0.1\\
$E_{\rm cut}$ (keV)	&13.9$^{+0.9}_{-0.2}$	&13.7$\pm$0.3	&18$\pm$2  	&18$\pm$2		&15$\pm$2\\
PL norm$^b$		&2.30$^{+0.09}_{-0.04}$	&3.6$^{+0.09}_{-0.06}$	&1.07$\pm$0.04	&1.34$\pm$0.08	&5.4$\pm$0.9\\
$E_{\rm Fe}$ (keV)	&6.44$\pm$0.08		&6.46$\pm$0.08	&6.8$\pm$0.2	&6.6$\pm$0.1		&6.4$\pm$0.2\\
$\sigma_{\rm Fe}$ (keV)	&0.5$\pm$0.2		&0.4$\pm$0.1	&0.6$^f$	&0.6$^f$		&0.4$^f$\\
EW(Fe) (eV)		&160$\pm$30		&120$\pm$20	&180$\pm$60	&160$\pm$50		&110$\pm$60\\
$E_{\rm edge}$ (keV)	&9.7$\pm$0.4		&9.7$\pm$0.4	&--		&--			&--\\
$\tau$			&0.05$\pm$0.01		&0.05$\pm$0.01	&--		&--			&--\\
Flux$^c$ (3-30 keV) 	&4.8			&8.0		&1.0		&1.5			&18.2\\
$\chi^2_r$/dof		&1.5/45			&1.2/45		&0.8/47		&0.6/47			&1.1/47\\
\hline \noalign{\smallskip}
\multicolumn{6}{c}{{\it INTEGRAL}/ISGRI + {\it RXTE}/PCA energy spectra$^{**}$ }\\
\hline \noalign{\smallskip}
Photon index		&--			&0.81$^f$	&--		&1.21$^f$		&0.7$^f$\\
$E_{\rm cut}$ (keV)	&--			&21$\pm$1	&--  		&19$\pm$4		&22$\pm$1\\
PL norm$^b$		&--			&1.5$\pm$0.9	&--		&0.2$\pm$0.2		&0.7$\pm$0.3\\
Flux$^c$ (2-100 keV)	&--			&10.7		&--		&1.9			&27.1\\
$\chi^2_r$/dof		&--			&1.1/63		&--		&0.7/62			&1.1/62\\
\hline \noalign{\smallskip}
\multicolumn{6}{c}{Power spectra}\\
\hline \noalign{\smallskip}
Power-law index		&1.7$\pm$0.1		&1.6$\pm$0.1	&2.1$\pm$0.4	&2.2$\pm$0.4		&1.39$\pm$0.05		\\
$rms$ (\%) (0.01-1 Hz)	&24$\pm$3		&25$\pm$3	&18$\pm$5	&18$\pm$4		&25$\pm$1\\
\hline \hline \noalign{\smallskip}
\multicolumn{6}{l}{$**$: It includes the model to the {\it RXTE}/PCA spectra
with the same value of the parameters} \\
\multicolumn{3}{l}{$a$: $\times 10^{22}$ cm$^{-2}$} &
\multicolumn{3}{l}{$c$: $\times 10^{-10}$ erg cm$^{-2}$ s$^{-1}$} \\
\multicolumn{3}{l}{$b$: $\times 10^{-2}$ cm$^{-2}$ s$^{-1}$ keV$^{-1}$ at 1 keV}&
\multicolumn{3}{l}{$f$: fixed}\\
\end{tabular}
\end{center}
\end{table*}

\subsection{X-ray states}

The long-term X-ray variability of \src\ is characterised by bright and
faint states (Fig.~\ref{asmpca}). The bright states appear as outbursts
that last typically for 15-20 orbital cycles, i.e. a few hundred of days.
These outbursts do not occur in a predictable manner but they do show a
general repeatable pattern with regard to the duration and profile of the
outburst. In general the outburst begins with a sharp flare that last for
one or two orbital periods. This flare is followed by a progressive
increase in the X-ray intensity until  a maximum is reached at about one
order of magnitude above the quiescent level. Assuming an absorbed
power-law with exponential cutoff (see below) and a distance of 6.8 kpc,
the X-ray luminosity at the peak of the outbursts ranges between $0.6-1.0
\times 10^{37}$ erg s$^{-1}$, while at the peak of the flare the luminosity
is typically a factor of 2 higher. During the bright state a modulation of
the X-ray intensity with orbital phase is clearly seen (Fig.~\ref{asmpca}),
but it is not detected in the faint state.  The average intensity
level of the faint state is $\sim$ 5 c s$^{-1}$ PCU$^{-1}$, which
corresponds to a X-ray luminosity of  $3.3 \times 10^{35}$ erg s$^{-1}$ in
the 3-30 keV range. Although the count rate may go down to $\sim$ 1 c
s$^{-1}$ PCU$^{-1}$ in the faint state, the source is detected in all PCA
pointings.  

In order to characterise the spectral and timing properties of the source
at different  states, we selected two subsets of PCA data corresponding to
the bright state and two to the faint state. Our selection was based on the
source activity, i.e. its PCA flux, and pulsar spin-up rate \citep[see Fig.
12 in][]{came07}. During the faint state the spin-up rate is close to zero
(or even negative),  whereas during the bright state it is $\simmore 2
\times 10^{-13}$ Hz s$^{-1}$. Thus, faint states correspond to epochs of
little or no mass accretion rate.  The duration (typically 8--10 orbital
periods) of the states was chosen such that it includes the rising and
declining phase of the bright state and provides good statistics for the
faint state. The time ranges of these intervals are given in Table 4 in Modified Julian Date
(MJD), below the state names. Additionally, we selected a one-day subset of
PCA data during the flare at MJD 54221 (Fig.~\ref{asmpca}). Note,
however, that RXTE/PCA observations were triggered by the sudden increase
of the \src\ flux in the ASM light curve. Therefore, the PCA observations
did not catch the peak of this flare but began when the flux was already in
the declining phase.  For each one of these states an average energy
spectrum was obtained. A systematic error of 0.6\% was added in quadrature
to the statistical error in the average energy spectra.

An absorbed power law with a high-energy cut off and an iron line at
$\sim6.4$ keV provided  good fits to all energy spectra in the range 3-30
keV.  The spectra of the bright state are distinctly harder than those of
the faint state, in agreement with previous findings \citep{bayk02,came07}.
The cutoff energy in the faint states occurs at slightly higher energies
than in the bright state, whereas the central energy of the iron line
is slightly larger in the faint state. The line width, however, is not
well constrained. If let as a free parameter then the line is broader in
the faint state. However, equally good fits are obtained in a relatively
wide range of values ($\sigma=0.3-1.0$ keV) if this parameter is fixed in
the faint state. While the addition of an edge at  $\sim$9 keV clearly
improves the fit in the bright state\footnote{The probability that the
improvement of the fit in the bright state occurs by chance by the addition
of this component is $\simless 10^{-3}$ based on the F-test.}, no such edge
is required for the faint state. The value of the best-fit parameters of
the flare spectrum agree with those of the bright state within the errors.

\begin{figure}
\resizebox{\hsize}{!}{\includegraphics{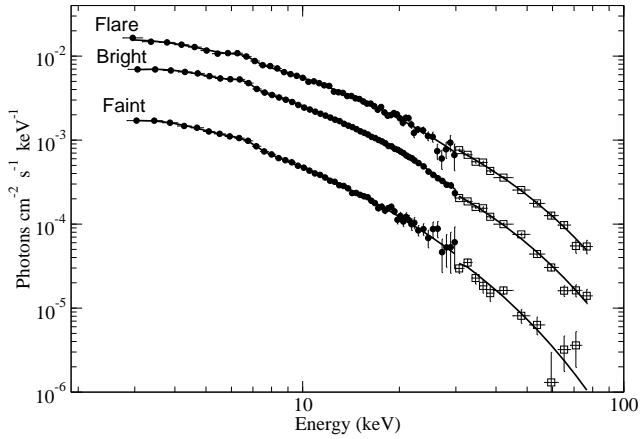} } 
\caption[]{Joint PCA+ISGRI spectra for the flare, bright2 and faint2 states. 
The model components comprises two cutoff power laws, a Gaussian and low-energy 
absorption. Circles correspond to PCA data and squares to ISGRI data.}
\label{spec}
\end{figure}

We also obtained an {\it INTEGRAL}/IBIS average energy spectrum for the
bright2 and faint2 states covering the energy range 20-100 keV. We selected
science windows where a statistically significant detection of \src\ was
found and in which the source lied within  12$^{\circ}$ off the centre.
This radius includes all observations in which the source was within the
PCFOV but it disregards the observations in which the source was too close
to the detector edge.  Figure~\ref{spec} shows the joint average PCA+ISGRI
spectra corresponding to the flare, bright and faint states. To fit these
spectra we used the same model as for the PCA data simply adding one extra
cutoff power law and a relative normalisation factor (fixed to 1 for the
PCA spectrum). However, we fixed the photon index of this new component to
the best-fit value found in the fit to the PCA data and allow only the
cutoff energy and normalisation to vary.  In the faint state  the addition
of {\it INTEGRAL} data to the PCA fit  represents a continuation of the
lower energy spectrum, that is, the two cutoff energies are consistent
within the errors. In contrast, the broad-band spectra  for the bright and
flare state are not well fitted by a single power law plus cutoff energy.
The extension of the spectra above 30 keV for these two states requires an
extra cutoff energy.   Table \ref{pcafit} shows the results of the spectral
fits. The parameters that fit the PCA spectra also fit the
PCA+INTEGRAL spectra. The only difference is the extra exponential decay
component.

\begin{figure}
\resizebox{\hsize}{!}{\includegraphics{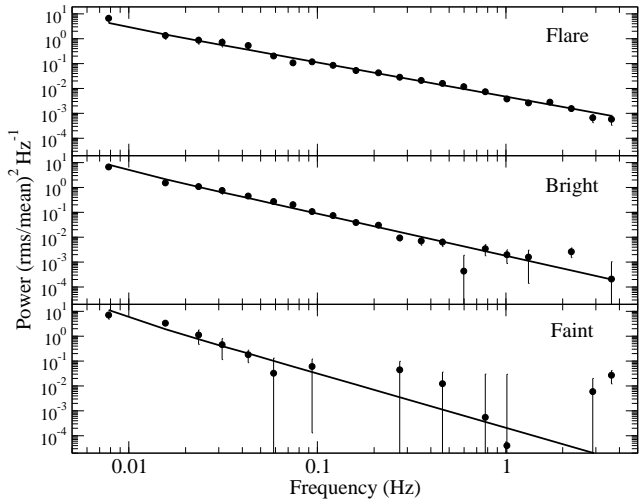} } 
\caption[]{Power spectra for the flare (MJD 54221), bright2 (MJD 52653) 
and faint2 (MJD 53265) states.}
\label{psd}
\end{figure}

The large effective area of the {\it RXTE} instruments make it especially
valuable for timing analysis of intensity variations from high-energy
sources. Therefore the study of the aperiodic variability of \src\ was
performed with PCA data only. Given the large number of observations and in
order to obtain an homogeneous set of power spectra in terms of energy and
frequency range coverage we used the configuration mode {\em Standard 1}.
The light curve corresponding to each PCA observation was binned with bin
size $0.125$ s, which is the maximum resolution provided by the {\em
Standard 1} data mode, and divided into segments of 128 s.  An FFT was
calculated for each segment giving power spectra covering the frequency
range 1/128--4 Hz. The final power spectrum for each observing interval
resulted after averaging all the individual power spectra and
logarithmically rebinning in frequency 
(Fig.~\ref{psd}). In general, the power spectra were well fitted to a
single power law.  The values of the power-law index and fractional
amplitude of variability ($rms$) given in Table~\ref{pcafit} correspond to
the mean and standard deviation  of the average values from all the
intervals included in each state. Given the intrinsic variability of the
source, we preferred to obtain mean values from the relevant observations
rather than one single average power spectrum from the entire light curve
of each state . Intervals resulting with less than six segments (i.e. six
power spectra) were not considered because of poor statistics (low number
of points in the low-frequency bins). Likewise, power spectra whose fits to
a power law gave reduced $\chi^2$ larger than 2 were rejected. Since the
flare state include one only observational interval the values quoted are
those from the average power spectrum of that interval. 
From the results shown in Table~\ref{pcafit} we conclude that the red noise
component becomes flatter as the X-ray flux increases.


\begin{figure}
\resizebox{\hsize}{!}{\includegraphics{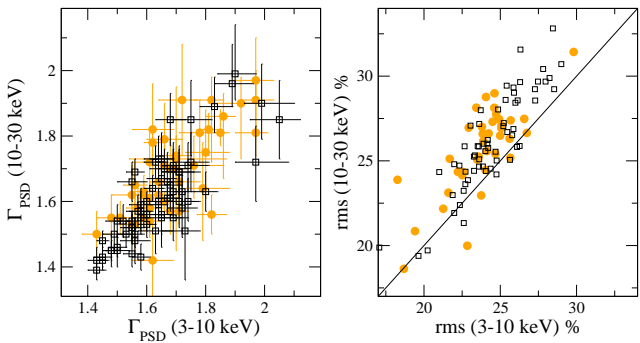} } 
\caption[]{Relationship between the power-law index and $rms$ resulting 
from the fits to the power spectra in two energy bands for the two bright 
states. Circles represent state 1 and squares correspond to state 2.}
\label{gsoft-ghard}
\end{figure}

\begin{figure}
\resizebox{\hsize}{!}{\includegraphics{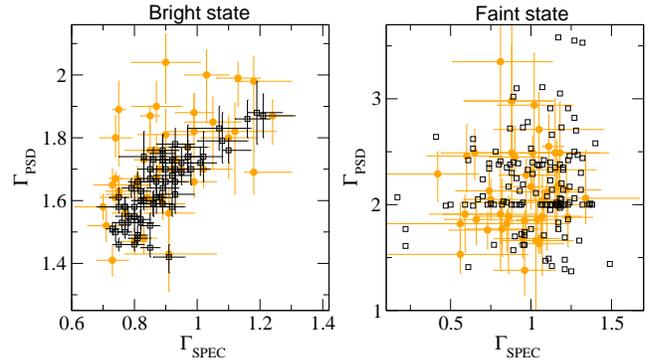} } 
\caption[]{The $\Gamma_{\rm SPEC}-\Gamma_{\rm PSD}$ diagram. Slope of the power spectral
continuum as a function of the photon index of the energy spectrum.
Circles represent state 1 and squares state 2. For clarity the 
error bars of the faint2 state were omitted.}
\label{gamma-gamma}
\end{figure}
\begin{figure}
\resizebox{\hsize}{!}{\includegraphics{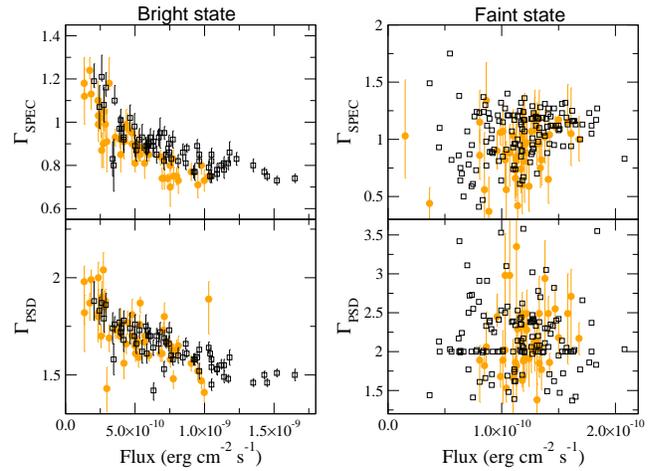} } 
\caption[]{Photon index and power-law index as a function of X-ray flux in 
the 3-30 keV range. Circles represent state 1, while squares correspond
to state 2. For clarity the error bars of the faint2 state were omitted.}
\label{parflux}
\end{figure}

\subsection{X-ray spectral-temporal correlation in the bright state}

In addition to characterising the bright and faint states we also searched
for correlations between the temporal and spectral parameters on shorter
timescales. For each PCA observation falling in one of the two
bright or two faint states defined above we obtained an energy spectrum and
two power spectra corresponding to the energy ranges 3--10 keV, 10--30 keV,
respectively. In this case the light curve at different energies were
extracted from data  of the {\em GoodXenon} configuration mode. As before,
the 3--30 keV energy spectra were fitted with an absorbed power law plus
exponential cutoff and a Gaussian representing fluorescence from cold iron,
while the power spectra exhibited strong red noise that was fitted with a
single power law. 

We first checked whether the aperiodic variability, i.e., the continuum of
the power spectra, depended on energy. Fig.~\ref{gsoft-ghard} shows the
power-law index obtained from the 3-10 keV and 10-30 keV power spectra for
the bright states. A fit of both sets of data to a straight line gives a
slope of 0.83$\pm$0.06, indicating that the power spectral continuum weakly
depends on energy. That is, on average, the slope of the power spectra at
higher energies is flatter than at lower energies. The fractional amplitude
of variability  in the frequency range 0.01--1 Hz is also, on average,
higher at high energies. As it can be seen in Fig.~\ref{gsoft-ghard}, the
$rms$ data points measured from the 10-30 keV power spectrum lie
systematically  above the one-to-one relationship, denoted by the straight
line.

Then we investigated the interdependency of the spectral and temporal
parameters.  Figure~\ref{gamma-gamma} shows the relationship between the
power-law index resulting from the power spectra using the 2--60 keV ({\em
Standard1}) light curves ($\Gamma_{\rm PSD}$) and the photon index from the
energy spectra ($\Gamma_{\rm SPEC}$). It is clear that in the bright state
steeper energy spectra correspond to steeper power spectra. However, this
correlation is lost in the faint state.

The variation of the temporal and spectral parameters as a function of the
3--30 keV flux is shown in Fig.~\ref{parflux}. Again correlations between
these parameters are apparent in the bright state only. The $\Gamma_{\rm
SPEC}$-flux plot confirms the known result that the source spectrum becomes
harder for higher flux \citep{bayk02,came07}. However, we find that this
correlation also holds in the timing domain with flatter power spectra
being associated with higher flux. The $\Gamma_{\rm SPEC}-\Gamma_{\rm PSD}$
(or $\Gamma_{\rm PSD}$-flux) represent new results, not previously reported
on a high-mass X-ray binary. 

\begin{figure}
\resizebox{\hsize}{!}{\includegraphics{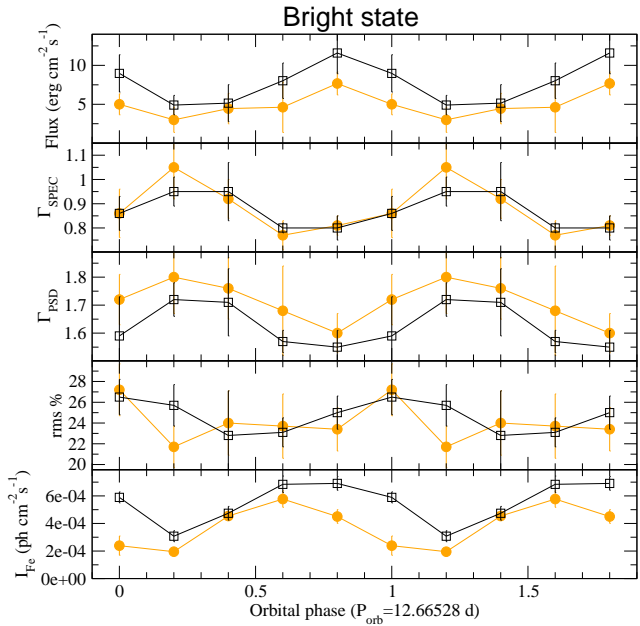} } 
\caption[]{Spectral and timing parameters as a function of orbital phase for the
bright state. Circles
correspond to state 1 and squares to state 2.}
\label{orbmod}
\end{figure}

\subsection{Orbital phase variability}

We also searched for orbital variability in the bright and faint states. We
assigned an orbital phase to each PCA observation falling within the four
intervals defined in Table~\ref{pcafit} using the orbital solution of
\citet{came07} (no significant difference was found in the results if the
orbital parameters of \citet{bayk07} were used). The data points were then
grouped into five phase bins. Figure \ref{orbmod} shows the results of the
orbital variability analysis. We show the bright state only as no
correlation with orbital phase was found in the faint state. The X-ray flux
in the 3-30 keV band in units of $10^{-10}$ erg cm$^{-2}$ s$^{-1}$, the
photon index of the energy spectra, $\Gamma_{\rm SPEC}$, the power-law
index of the power spectra, $\Gamma_{\rm PSD}$, the fractional amplitude of
variability in the frequency range 0.01--1 Hz, $rms$, and the intensity of
the iron line, $I_{\rm Fe}$, are shown as a function of orbital phase. In
this figure, periastron corresponds to orbital phase $\sim$0.4, while
maximum flux is reached around phase 0.7. 

When the 3-30 keV flux is high, the intensity of the iron line is also
high, while the photon index and power-law index are low. The $rms$ also
shows a weak modulation: the source appears more variable when the flux is
high, but the relatively large errors diminish the significance of this trend.
The iron line energy remains fairly constant throughout the orbit. A
weighted mean gives 6.44$\pm$0.06 keV. 

\section{Discussion}

\subsection{Optical analysis}

Numerous studies on Be stars
\citep{dach86,humm95,hanu96,port03,tycn05,jone08} showed that the
equatorial disc in Be stars is rotationally supported consistent with a
Keplerian velocity field and that the strength and shape of the \ha\ line
provides important information on the structure of the disc. Double-peaked
profiles have been interpreted as a result of the Keplerian motion of the
particles in the disc, with the peak separation giving an estimate of the
rotation velocity. Split profiles are expected to occur when the disc is
smaller, that is, at smaller radii. When the disc grows, since the
Keplerian velocity is inversely proportional to the radius, the velocity
becomes too small to be discernible in low or intermediate resolution
spectra. This trend of double-peaked profiles merging in to one single-peak
line as the H$\alpha$ equivalent width increases is a well known result
both in Be stars \citep{dach86} and Be/X \citep{reig00}.

We find a good correlation between the strength and shape of the \ha\ line
and the X-ray intensity in \src\ (Fig.~\ref{xopt}). Emission line profiles
are seen only when the source is bright in X-rays, while low-intensity
X-ray states are associated with absorption line profiles. This type of
correlation has been seen in other Be/X. What makes \src\ unique is the
rapidity of the changes in the strength and shape of the \ha\ line,
presumably due to a the rapid growth of the disc and the almost immediate
transfer of matter into the neutron star to generate X rays. Typically, the
time scales for the formation and disappearance of the circumstellar disc
in BeX are larger than 3 years \citep[see e.g.][]{reig05}. In \src\ it is 1--2
years. Thanks to our monitoring in the optical band we can put some
constraints on the duration of the formation and subsequent loss of the
disc (see Fig.~\ref{xewR} and Table~\ref{ewha}). The \ha\ line was in
absorption with \ew=+2.4 \AA\ without any trace of fill-in emission in June
2006. The largest value of the \ew, which would roughly correspond to the
maximum extension of the circumstellar disc, was achieved in September
2007, that is, \src\ took about 14 months to develop the circumstellar
disk. This duration should be considered as an upper limit. Given the drop
in X-ray intensity around January 2007 and the lack of observations in the
optical band at this time, it is very likely that the \ha\ was in
absorption at the beginning of 2007. If this is the case then the formation
of the disk took just about 8 months. By December 2007, the strength of the
\ha\ line was clearly in the decline, turning into an absorption phase
(\ew\  had weakened to just 1 \AA, although still  in emission). Full
absorption at a level of \ew=+2.3 \AA\ was seen about one year after
maximum. We conclude that the total duration of the formation/dissipation
of the circumstellar disc in \src\ is $\simless$ 2 yr, and most probably of
the order of 1.3--1.5 yr. This duration compares to 3-5 yr of 4U\,0115+63
\citep{reig07}, 4-5 yr of  A\,0535+26 \citep{haig04} and V0332+53
\citep{gora01}, 4 yr of LS\,992 \citep{reig01} and 7 yr of X-Per
\citep{clar01}. The reason for these fast changes must lie in the
relatively narrow orbit (\src\ is the Be/X with the shortest orbital
period). 

In \src\ changes in the disc happen so fast that the system does not have
the time to develop a large disc. The fact that {\em i)} the largest \ew\
measured in \src\ is $\sim$ -5 \AA, a relatively small value and {\em ii)}
the emission line profile appears always as double peaked, provide evidence
in support of a small disc radius. The narrow orbit and continuous passages
of the neutron star limit the growth of the disc and do not allow the Be
star to develop an extended disc. On the other hand, the duration of the
disc-loss phase also seems to be short. In fact, it is likely that the
complete loss of the disc never materialises. An inspection of
Table~\ref{ewha} reveals that CQE and emission profiles are present in  the
majority of the observations. As stated above emission lines are formed in
the  disc. CQE profiles also require that a circumstellar disc exists,
although of a much smaller spatial extent \citep{rivi99}. Note also that
some of the profiles classed as pure absorption may contain a weak central
peak that is not apparent due to the low spectral resolution. Another
interesting characteristic of the X-ray behaviour of \src\ is the large and
sudden increase in intensity that precedes some of the outbursts. The 
X-ray intensity increases by a factor of $\sim$20 in a few days. The onset
of these flares occurs at or near periastron  and last for no more than two
or three orbital cycles. Then the X-ray intensity abruptly falls to almost
pre-flare values. Immediately after, a lower intensity and longer outburst
begins (the bright state).

Thus the picture that emerges to explain the long-term X-ray/optical
variability is the following: the Be star's equatorial disc forms due to
ejections of matter from the Be star's photosphere through a still unknown
mechanism. As the disc grows the \ha\ line turns from absorption into
emission. The fact that the emission line profile is always split, i.e., no
single-peak line is seen and the relatively small \ew\ indicates a small
disc.  We can estimate the disc radius from the separation of the \ha\
profile peaks, as this value and the outer radius ($R_{ \rm disc}$) of the
emission line forming region are related by \citep[see e.g.][]{humm95}

\begin{equation} 
\frac{R_{\rm disc}}{R_*}=\left(\frac{2 v \sin i}{\Delta_{\rm peak}}\right)^2 
\end{equation}

\noindent where $v\sin i$ is the projected rotational velocity of the B
star ($v$ is the equatorial rotational velocity and $i$ the inclination
toward the observer). For \src\ $v\sin i\approx 240$ km s$^{-1}$
\citep{reig04}, while the
average peak separation, as measured from the September
2007 spectra is $\Delta_{\rm peak}=340$ km s$^{-1}$. Thus R$_{\rm
disc} \sim 2-3 $ R$_*$.
Taking the characteristic values of mass and radius of a B0 star,
namely $M_*=20 \msun$ and $R_*=8 \rsun$ for the B star \citep{vacc96} and
the canonical mass of a neutron star $M_x=1.4 \msun$, the orbital period
($P_{\rm orb}=12.6$ d) implies, according to Kepler's third law, an orbital
separation of $a\approx 63 \rsun$ or equivalently $a\approx 8 $ $R_*$.
Correspondingly, periastron lies at $R_{\rm per}=4.7$ $R_*$ for an
eccentricity $e=0.4$. Similarly, the radius of the Roche lobe of the
primary star is \citep{pacz71} $R_{RL}/a=0.38+0.20\log(M_*/M_x)=0.6$. That
is, the radius of the disc is similar to the size of the Roche lobe, which
in turn, is similar to the periastron distance,
$R_{RL}\approx R_{\rm per} \sim R_{\rm disc}$. 

We conclude that due to the small orbit, the neutron star truncates the
disc at very small radii. It is very likely that, occasionally, the neutron
star physically impinges on the disc during periastron passage giving rise
to sharp and intense X-ray flares. Note that the fact that $R_{\rm disc}=2
\, R_*$ does not mean that the disc terminates abruptly at that distance,
but matter would extend to larger radii following a density law of the form
$\rho \sim r^{-\beta}$, where $\beta\approx 2.5-3$ \citep{wate86,wate88}.
The fact that the optical indicators (magnitudes and \ew) continue rising
after the flare episode implies that the disc has not yet been destroyed.
In subsequent periastron passages, matter is still transferred to the
neutron star but in a less dramatic way. An accretion disc around the
neutron star is likely to be formed, as indicated by the spin-up episodes
\citep{inam04, came07}. When the fuel that power the X-ray activity is used
up, that is to say, when the circumstellar disc disappears then accretion
ceases and the source returns to an X-ray quiescent state. 

\subsection{X-ray analysis}

Our X-ray spectral analysis of the bright and faint states of \src\ confirms previous
reports \citep{bayk02,bayk07,came07} that in the bright state the spectrum is
harder than in the faint state. This result is opposite to what it is seen in
other types of X-ray binaries but it is not exclusive of \src. Black-hole
systems and low-mass X-ray binaries display different source states \citep[][and
references therein]{klis06}. In black-hole systems three basic states are
identified, namely, soft, hard and intermediate (which includes the so called
very high state). Although these states may occur at any luminosity, when
the black-hole system is followed through an outburst, the soft state normally
corresponds to higher X-ray intensity values, while the hard state is normally
seen at the beginning and end of the outburst, i.e, at lower count rates. 
Likewise, spectrally hard states in low-mass X-ray binaries occur at lower
luminosities than softer spectra states \citep{oliv03,macc03}. In contrast, in a spectral
and timing study of four Be/X-ray during major outbursts, \citet{reig08} found
that three of the sources investigated showed low/soft spectral states (the
so-called horizontal branch) at the beginning and at the end of the outburst.

The origin of these differences must lie on the accretion processes in the
vicinity of the compact object. Unlike black-hole systems and low-mass
X-ray binaries with weak magnetic-field neutron stars, the compact object
in Be/X-ray binaries  is a strongly magnetised neutron star,  which results
in significant coupling of the magnetic field and the accretion flow 
\citep{whit83}. One of the results of this coupling is that the accretion
flow  is threaded onto the neutron star magnetic field lines and channeled
on to the magnetic poles, producing two or more localised X-ray hot spots 
\citep[see e.g][]{fran02}.  Above these hot spots the magnetic field lines
adopt the configuration of a roughly column-shaped surface or funnel, which
is referred to as the accretion column, the polar cap being its base. The
polar caps are treated as dense thermal mounds with a blackbody spectrum.
The detection of such blackbody components at low energies in a number of
low-luminosity X-ray pulsars give support to the existence of such thermal
mounds \citep{cobu01,mukh05,palo06,reig09}. The surface of the mound  is
responsible for photon creation and absorption \citep{beck05}. The
high-energy emission observed from accreting pulsars results from the
formation of a standing shock wave on this accretion column:  the energetic
particles in the accretion flow comptonise  the thermal emission produced
on the polar cap \citep{beck05}.  Increased mass accretion rates are
expected to result  in harder X-ray  spectra \citep{lang82,beck05} because
the photons spend more time, on average, being upscattered in the flow
before escaping.

The faint state corresponds to a quiescence and stable state of the source
in which the X-rays are no longer powered by disc accretion but most likely
are the result of wind accretion.  Wind-fed system, like supergiant X-ray
binaries, show erratic and flaring (high-amplitude changes on timescales of
seconds) X-ray variability that it is consequence of inhomogeneities in the
accretion wind. This type of variability is almost absent in \src, where the
main source of variability at short timescales are the X-ray pulsations.
The fact that we do not see random high-amplitude variations during the
faint state of \src\ might indicate that the dominant accretion wind is the
more stable equatorial low-velocity high-density wind characteristic of BeX
\citep{wate88}. This wind, that ultimately would create the circumstellar
disc, would be highly suppressed during the faint state. It would also
indicate that the neutron star orbit is coplanar to the plane of the
circumstellar disc. Otherwise, the neutron star would be exposed to the
higher velocity and more inhomogeneous polar winds, typical of massive
stars, giving rise to erratic X-ray variability. The narrow orbit of \src\
would also help avoid the polar winds if the orbit is coplanar to the disc.
Occasionally sudden and short-lived increases in X-ray flux are seen
superimposed on a long-term decline phase during the faint state (see e.g.
MJD 52800--53200 in Fig.~\ref{asmpca}). These mini-outbursts may correspond
to fail attempts to create the circumstellar disc, i.e. ejection of matter
from the Be star's photosphere that are not retained by the star.

In contrast, the bright state exhibits high richness in variability. Both
spectral and temporal parameters correlate with one another and with X-ray
flux and orbital phase. During the orbital maximum the spectral photon
index is minimum. Similar variability behaviour has been reported in the
supergiant X-ray binary 2S 0114+650 \citep{farr08}. Based on the
similarities with the eclipsing X-ray pulsar EXO 1722--363 \citep{thom07},
\citet{farr08} attributed the spectral variability to absorption effects as
the neutron star passes behind a heavily absorbing wind.

We have found for the first time in a HMXB a correlation between the shape
of the energy and power spectra, namely softer energy spectra correspond to
the disappearance of power at high frequencies, i.e. softer power spectra.
Although there are no detailed models of the power spectra arising from
accretion onto highly magnetized neutron stars, one would expect that the
simple scenario described above would also explain the $\Gamma_{\rm
SPEC}-\Gamma_{\rm PSD}$ correlation (Fig.~\ref{gamma-gamma}) At high mass
accretion rates, i.e., when the X-ray flux is high, the magnetosphere
shrinks, as the magnetosphere's radius $r_{\rm m}\propto
\dot{M}^{-2/7}$ \citep[see e.g.][]{davi73}. If the source of variability in the power
spectrum comes from the interaction between the accretion disc and the
magnetosphere then smaller magnetospheric radius implies shorter
characteristic time scales (or higher frequencies), hence flatter power
spectra. When the flux decreases (softer energy spectrum), the size of the
magnetosphere increases and so does the characteristic timescale of
variability producing more power at low frequencies, hence steeper power
spectra. The fact that the $rms$ is similar in the bright and faint states
supports the notion that the source of variability is similar in both
states and the only difference is the size of the emitting region. 

If we assume that the characteristic frequencies result from Keplerian
motion in the inner parts of the accretion disc, and that the accretion
disc extends all the way down to the magnetosphere, then the characteristic
Keplerian frequency of an orbiting free particle is 

\[\nu=\sqrt{\frac{GM}{4\pi^2r_{\rm m}^3}} \]

\noindent Assuming that all the kinetic energy of infalling matter is given
up to radiation, i.e., $L_{\rm x}=GM_{\rm x}\dot{M}/R_{\rm x}$, an increase
of the X-ray luminosity by a factor of $\sim$6, as observed in the bright
state of \src, implies a reduction in the size of the magnetosphere by
$\sim40$\%, while the characteristic frequency of variability increases by
a factor $\sim$2.

In Fig.~\ref{gsoft-ghard} we showed that the fractional $rms$ is larger at
higher energy. If the source of variability were variations in the soft
photon input, then simple Comptonization models would predict a decrease of
the amplitude of variability with increasing photon energy. This trend is
the result of the averaging effect of Compton scatterings: low-energy
photons  retain most of the variability of the seed photon input because
they have not suffered many scatterings. In contrast, the variability of
high-energy photons is smeared out as these photons spend longer time in
the comptonizing medium. These prediction are not consistent with
observations of black-hole systems. In low-mass X-ray binaries and
black-hole systems, the fractional $rms$ of discrete noise components (i.e.
QPO) tend to increase with photon energy \citep{gilf03,homa01}. In the case
of broad-band noise, the relationship of the $rms$ with energy shows a much
more complex behaviour \citep{lin00}, and it is source state dependent
\citep{gier05}. Note also that in black-hole binaries the geometry of the
Comptonizing medium is very different from the geometry in the pulsar
accretion column and the optical depth is expected to be lower. In order to
have larger fractional $rms$ at higher energies, as shown in
Fig~\ref{gsoft-ghard}, the source of variability cannot be solely due to
intrinsic variations of the source of soft photons. \citet{gier05} found
that an increase in the fractional $rms$ with photon energy can be achieved
if the power released in the comptonising plasma changes, i.e., if the
energy of the comptonising electrons changes (electron acceleration and/or
direct heating). Applied to \src\, this would mean that the power supplied
to the electrons in the accretion flow changes, perhaps as a result of
varying mass accretion rate, which may well affect the geometry and optical
depth of the Comptonizing column. 

\section{Conclusions}

We have performed an analysis of the X-ray timing and spectral properties
of the Be/X-ray binary \src\ in correlation with optical spectroscopic
observations of the \ha\ line. We find a good correlation between the
strength and profile of this line and the X-ray activity of the source.
Emission line profiles are observed during bright X-ray states. In these
states the \ha\ equivalent width is largest. During faint X-ray states \ha\
appears in absorption. \src\ exhibits the fastest time scales for the
disappearance and reformation of the circumstellar disc, compared to other
typical Be/X-ray binaries. In less than 2 years \src\ is capable to form
and lose the disc. Due to its narrow orbit, the neutron star prevents the
disc from extending to large radii; the disc is truncated at 2--3
stellar radii. Based on the \ha\ line profile we set stringent constraints
on the size of the circumstellar disc and we find that its size can reach
the Roche-lobe radius of the system (which it is similar to the periastron
distance of the two objects).

We have also performed for the first time in a Be/X-ray binary a correlated
study of the X-ray spectral and timing properties and found a correlation
between the shape of the energy spectrum and that of the power spectrum in
the bright state. When the energy spectrum is soft, the power at high
frequencies is suppressed. This correlation, together with the existence of
hard/high and soft/low states, has been interpreted in the context of
emission from the accretion column.

\section*{Acknowledgments}

We thank all the observers that helped obtain the optical spectra: P.
Berlind and M. Calkins from FLW observatory and  A. Manousakis, M. Lanzara,
E. Beklen, and E. Nespoli from the SKI observatory. This work has been 
supported in part by the European Union Marie Curie grant
MTKD-CT-2006-039965 and EU FP7 "Capacities" GA No206469.  A. S\l{}owikowska
is partially supported by the Polish Ministry of Science and Higher
Education project 362/1/N-INTEGRAL (2009-2012). This work has made use of
NASA's Astrophysics Data System Bibliographic Services and of the SIMBAD
database, operated at the CDS, Strasbourg, France. The ASM light curve was
obtained from the quick-look results provided by the {\it RXTE}/ASM team.
{\it SWIFT}/BAT transient monitor results provided by the {\it SWIFT}/BAT
team. Skinakas Observatory is a collaborative project of the University of
Crete, the Foundation for Research and Technology-Hellas and the
Max-Planck-Institut f\"ur Extraterrestrische Physik.

\label{lastpage}

\end{document}